\documentstyle[12pt]{article}
\def\a{\alpha}
\def\b{\beta}
\def\g{\gamma}
\def\d{\delta}
\def\e{\epsilon}
\def\f{\phi}
\def\F{\Phi}
\def\z{\zeta}
\def\y{\eta}

\def\k{\kappa}
\def\l{\lambda}
\def\m{\mu}
\def\n{\nu}
\def\r{\rho}
\def\p{\pi}
\def\s{\sigma}

\def\t{\tau}
\def\v{\varphi}
\def\o{\omega}

\def\L{\Lambda}
\def\part{\partial}

\def\be{\begin{equation}}
\def\ee{\end{equation}}

\newcommand{\bfp}{\mbox{\boldmath$\p$\unboldmath}}
\newcommand{\bfy}{\mbox{\boldmath$\y$\unboldmath}}
\newcommand{\bfpsi}{\mbox{\boldmath$\psi$\unboldmath}}

\begin{document}

\title{\Large Reparametrization Invariance  as Gauge Symmetry}
\author{G. F\"ul\"op, D.M. Gitman\\
{\small Instituto de F\'{\i}sica, Universidade de S\~ao Paulo}\\
{\small P.O. Box 66318, 05315-970 S\~ao Paulo, SP, Brasil}\\
I.V. Tyutin\\
{\small Lebedev Physical Institute, 117924 Moscow, Russia}}

\maketitle

{\vskip -10cm 
\noindent S\~ao Paulo - Moscow\\
 1997\\
IFUSP/P-1263\\
hep-th/9805040}

\vskip 10cm
\begin{abstract}
Reparametrization invariance being treated
as a gauge symmetry shows some specific peculiarities.
We study these peculiarities  both from a general
point of view and on concrete examples. We consider the canonical
treatment of reparametrization
invariant systems in which one fixes the gauge on the
classical level by means of time-dependent gauge conditions.
In such an approach one can interpret different gauges
as different reference frames.
We discuss the relations between different gauges and the problem
of gauge invariance in this case. Finally, we establish a general
structure of reparametrizations and its connection with the
zero-Hamiltonian
phenomenon.
\end{abstract}

\setcounter{page}{1}
\setcounter{equation}{0}

\section{Introduction}                       \label{bevezetes}

Many  actual physical theories are formulated in the
 so called reparametrization invariant
(RI) form, for instance, models of point-like relativistic particles,
gravity and string theory.
Formally, reparametrization invariance
can be treated as a gauge symmetry.
However, this gauge symmetry shows some peculiarities, so that
it is natural to separate it in a special class of gauge symmetries.
Due to the same reason one has to be  careful when formally
applying recipes extracted from the consideration of
gauge symmetries of different nature.
In all known examples of RI theories the Hamiltonian
vanishes on the constraint surface,
in spite of the fact that  explicit forms of
the reparametrization transformations in these examples may look
different.
This issue raises a question:
what is the general structure of such transformations
and is there a definite relation between such a
structure and  zero-Hamiltonian phenomenon?
The zero-Hamiltonian phenomenon in RI theories
raises another well-known problem: what is time-evolution
in this case?
This question has a principal character for the
construction of an adequate quantum theory of gravity.
In the canonical schemes of consideration there exists a
possibility to introduce the evolution by means of a
time-dependent gauge fixing.
In turn, this demands a modification of the standard scheme of canonical
quantization \cite{ab1}, which is adopted to stationary second-class
constraints
(such a  modification  was
first proposed in \cite{g-t}).
Fixing the gauge in such a manner we get
different evolutions depending on the selected gauge.
And here we meet the question well-known in gauge theories:
to what extent does the physical content of a theory
depend on the gauge fixing and what is gauge invariance here? There
exist, in fact, two essentially different points of view on this problem.
According to the first one, which is called "local" point of view, the gauge
fixing of the reparametrization gauge freedom corresponds to a certain choice
of the reference frame (RF). At the same time space-time variables in
the Lagrangian have to be identified namely with the coordinates of the above
RF. The reparametrizations relate the description 
of the system in different RF.
Thus, one has to admit that local physical quantities may depend on the
choice of the gauge. Another, "non-local" point of view, assumes that there
exists a reparametrization invariant  description.
Supporters of this position believe that such a description may be realized
if one includes an observer in the frame of the theory. Then the physical
quantities do not depend on the choice of the gauge, which fixes the
reparametrization freedom, and must commute with the corresponding
first-class constraints. Unfortunately, the "non-local" point of view
remains, in the main, declarative. It seems that its clear and convincing
realization is absent until now. An exelent and detailed survey on the
subject (and relevant references) one can find in \cite{isham}.

In the present paper we discuss the above and some other questions
related to RI theories both from a general point of view and
on specific examples. We advocate the "local" point of view considering
several examples, where one can compare RI and non-RI versions of the same
theory. Namely, we study a finite-dimensional theory, a field theory in a
flat space-time, and a theory of the relativistic particle, all of them both
in non-RI and RI form. We remind briefly on the treatment of systems with
non-stationary second-class constraints and apply then this formalism to the
above mentioned theories to impose time dependent (space-time dependent)
gauges. We analyse the 
relation between different gauges both on the classical and the
quantum level. Based on the considered examples we formulate an
interpretation which, in fact, supports the "local" point of view. We argue
that the reparametrization symmetry has to be treated specially from the
gauge symmetries of different nature. In the final part of the paper, which
looks more formal and independent from the previous conceptual part, we
study the general structure of the reparametrizations and its relation with
zero Hamiltonian phenomenon. On the
example of the time reparametrization we propose a general definition of
reparametrization symmetry transformations.

\setcounter{equation}{0}
\section{Introducing Reparametrization Invariance}

The action  of a point-like relativistic
particle
\be
S =  \int_0^1 L d \t~, \;  
L = - m \sqrt{\dot{x}^2}~, \; 
x = (x^{\m})~, \; 
  \dot{x}^{\m} = \frac{dx^{\m}}{d\t}~, \;
\m = 0,...,D~,                                       \label{action1}
\ee
gives us a simple example of RI theory.
 It is invariant under reparametrizations
$x^{\m}(\t) \rightarrow x'^{\m}(\t) = x^{\m}(f(\t))$,
where $f$ is an arbitrary
function obeying only the following demands:
$\dot{f}(\t) > 
0, f(0) = 0, f(1) =1$.
The reparametrizations  can be interpreted as  gauge transformations
(GT) whose infinitesimal  form is
\be
\d x^{\m}(\t) = \dot{x}^{\m}(\t) \e(\t)~,   \hskip 5mm      \label{II2}
\d L = \frac{d}{d \t} [ \e(t) L ]~,
\ee
where $\e(\t)$ is a time dependent parameter.
An equivalent Lagrangian function,
which is adapted to the $m\rightarrow 0 $
limit, contains an additional variable
$e$ and is of the form:
\be
L  =  -\frac{\dot{x}^2}{2e} - e \frac{m^2}{2}~.
                                                            \label{II3}
\ee
Here the infinitesimal form of the reparametrizations is:
\be
\d x^{\m}(\t) = \dot{x}^{\m}(\t) \e(\t)~,
\hskip 5mm \d e(\t) = \dot{e}(\t) \e(\t) + e(\t) \dot{\e}(\t)~,
\hskip 5mm
\d L = \frac{d}{d\t} [\e(\t)L]~. \label{II4}
\ee
String theory is of the same nature,
its action is invariant
under the reparametrizations of two variables.
Gravity is an example of RI
 field theory.
The Einstein action
\be
S_E = \int {\cal L}_E d^{D+1}x~, \hskip 5mm
{\cal L}_E = - \sqrt{-g}R,
\ee
is invariant under general coordinate transformations
$x^{\m}\rightarrow x'^{\m}=x'^{\m}(x)$,
$g_{\m\n}(x) \rightarrow g'_{\m\n}(x)$,
$g'_{\m\n}(x')=
\frac{\part x^{\l}}{\part x'^{\m}} \frac{\part x^{\s}}{\part x'^{\n}}
g_{\l\s}(x)$.
These are, in fact, reparametrizations of $D+1$ space-time variables.
They can be treated as GT,
\be
\d g_{\m\n}(x) = D_{\m} \e_{\n}(x) + D_{\n} \e_{\m}(x)~, \hskip 5mm
\d {\cal L}_E = \part_{\m} [\e^{\m}(x) {\cal L}_E]~,     \label{deltag}
\ee
where 
$\e^{\m}(x)$
are GT parameters - arbitrary functions on space-time coordinates.

Any action can be extended to a RI
form \cite{bergman}.
Consider, for example, a non-singular action
(similar consideration can be made for any singular
Lagrangian as well)
\be
S = \int^{t_2}_{t_1} L({\bf{x}}, \dot{\bf{x}}, t) dt,
\hskip 5mm
{\bf{x}} = (x^i),~ i=1,...,D~,
\hskip 5mm
\dot{\bf{x}} = \frac{d {\bf{x}}}{dt}~.             \label{m1}
\ee
Let us change $t$ to $x^0$ and then  replace
 the integration variable $x^0$,
\be
x^0 = f(t), \hskip 5mm f(t_1) = t_1 \hskip 5mm
f(t_2) = t_2 ~.                                         \label{m2}
\ee
Thus, we get
\be
S_R = \int^{t_2}_{t_1} L_R(x, \dot{x}) dt~,              \label{m4}
\hskip 5mm  L_R(x, \dot{x}) =  L({\bf{x}},
\frac{\dot{\bf{x}}}{\dot{x}^0}, x^0) \dot{x}^0~.       
\ee
As long as we keep in mind the relations (\ref{m2}),
the action (\ref{m4}) is completely equivalent to the initial one (\ref{m1}).
On the other hand, one can now treat (\ref{m4}) in a
new way, namely, one can forget about (\ref{m2}) and
treat $x^0$ as a new independent variable, so that
the total set of variables of the theory is
$x = (x^{\m})=  (x^0, \bf{x})$.

Let us analyse the relation between the theory with the action
(\ref{m4}) and (\ref{m1}), in particular, in the Hamiltonian
formulation.
For the non-singular theory (\ref{m1})
one can always solve
the equations, which define the momenta, with respect to all
velocities:
\be
{\bfp} = \frac{\part L}{\part \dot{\bf x}}
\Rightarrow \dot{\bf x} = {\bfpsi}({\bf{x}}, {\bfp},t),
\hskip 5mm {\bfp} = (\p_i) ~.                               \label{m8}
\ee
Then the time evolution is generated by the Hamiltonian
equations without any constraints,
\be
\dot{{\bfy}} = \{ {\bfy}, H\}~,
\hskip 5mm {\bfy} = ({\bf x, \bfp}) ~,                 
\hskip 5mm
H = \left.\left( \frac{\part L}{\part \dot{\bf x}}
\dot{\bf x} - L \right)\right|_{\dot{\bf{x}} = {\bfpsi}}
= H({\bf{x}}, {\bfp}, t)~.                                 \label{m10}
\ee
In the theory with the action $S_R$ there appear primary
constraints in the Hamiltonian formulation.
Indeed, let $\p_{\m} = (\p_0, \bfp)$ be momenta conjugated to $x^{\m}$,
\be 
 \p_0 = \frac{\part L_R}{\part \dot{x}^0} =
- \left.\left( \frac{\part L}{\part \dot{\bf x}}\dot{\bf x} -
L\right)\right|_{\dot{\bf{x}} \rightarrow
\frac{{\dot{\bf{x}}}}{\dot{\bf{x}}^0}, t \rightarrow x^0}~, \hskip 5mm
\bfp = \frac{\part L_R}{\part \dot{\bf x}} =
\left. \frac{\part L}{\part \dot{\bf x}}
\right|_{\dot{\bf{x}} \rightarrow
\frac{{\dot{\bf{x}}}}{{\dot{x}^0}}, t \rightarrow x^0}~.
\label{m12}
\ee 
>From the second equation in (\ref{m12}) (taking into account (\ref{m8}))
we get:
$
{\dot{\bf x}} = {{\dot{x}^0}}  \bfpsi({\bf{x}},
{\bfp},x^0),$
whereas ${\dot{x}}^0$ is a primarily unexpressible velocity.
Then the first equation  (\ref{m12}) (taking into account (\ref{m10}))
appears to be a primary constraint
\be
\f_1 =  \p_0 +  H({\bf{x}}, {\bfp},x^0) =0~.   \label{m13}
\ee
Constructing the total Hamiltonian $H^{(1)}$ according to the standart
procedure \cite{ab1,g-t} we get \be H^{(1)} = \left.\left( \frac{\part
L_R}{\part \dot{x}^{\m}}\dot{x}^{\m} -L_R \right)\right|_{\dot{\bf x} =
\dot{x}^0 \bfpsi({\bf{x}}, {\bfp}, x^0)} = \l \f_{1}, \hskip 5mm \l =
\dot{x}^0~.  \label{m14} \ee Thus, the total Hamiltonian  vanishes on the
constraint surface (on the equations of motion).  No more constraints appear
from the consistency conditions.  To fix a gauge we have to impose a new
constraint $\f_2  =0$, so that the matrix $\{ \f_a, \f_b \},~ a,b =1,2 $ is
not singular.  A natural form of a such a condition is $ \f_2 = x^0 -
\v({\bf{x}},{ \bfp}, t) =0~,          
\label{m15} $ where the function
$\v({\bf{x}}, {\bfp}, t)$ has an essential 
$t$-dependence,
introduced in the theory, in spite of the fact, 
that the Hamiltonian is zero.
The simplest choice of the gauge  condition is
(we will call such a condition - {\it chronological gauge}),
\be
\f_2 = x^0 - t =0~.                                     \label{m16}
\ee
The set of second-class constraints (\ref{m13}), (\ref{m16})
explicitly depends on time.
The general method to deal with non-stationary
constraints in the canonical formulation and quantization procedure
were first proposed in
\cite{g-t}. Then similar results were obtained by a
 geometrical approach in \cite{evans1}. The BRST formulation
 of the non-stationary constraints case was discussed in \cite{batalin}.
Below we  briefly remind on the treatment \cite{g-t} of systems with
non-stationary second--class constraints.

Consider a theory with second-class constraints
$\phi_a(\eta,t)=0$ (where $\eta=(x^i,\pi_i)$ are canonical variables)
which may explicitly depend on time $t$. Then
the equation of motion of such a system may be written
by means of the Dirac brackets, if one
 formally introduces a momentum  $\epsilon$
conjugated to the time $t$, and
defines the Poisson bracket in the extended
phase space of canonical variables
$(\eta;t,\epsilon)$,

\begin{equation}
\dot{\eta}=\{\eta,H + \epsilon\}_{D(\phi)},\;  \phi(\eta, t) =0~,
\label{e1}
\end{equation}

\noindent where H is the Hamiltonian of the system, and
$\{A,B\}_{D(\phi)}$
is the notation
for the Dirac bracket with respect to the system of  second--class
constraints $\phi$. The Poisson brackets, wherever encountered,
are henceforth
understood as one in the above mentioned extended space.
The quantization procedure in ``quasi-Schr\"odinger''
 picture can be formulated in
that case as follows. The variables  $\eta$   of the theory are
assigned the
operators  $ \tilde{\eta} $,  which satisfy the
following relations
\begin{equation}\label{e2}
[\tilde{\eta},\tilde{\eta}']=i\{\eta,\eta'\}_{D(\phi)}|_{\eta=
\tilde{\eta}}~, \hskip 5mm
\phi(\tilde{\eta},t)=0   , \label{e3}
\end{equation}
and equations of evolution
\begin{eqnarray}
\dot{\tilde{\eta}}
&=&\{\eta,\epsilon\}_{D(\phi)}|_{\eta=\tilde{\eta}} 
= - \left.\{\eta,\phi_{a}\} C_{ab}\frac{\partial
\phi_{b}}{\partial t}\right|_{\eta=\tilde{\eta}}~, \hskip 5mm 
C_{ac} \{\phi_{c},\phi_{b} \} = \d_{ab}~  .\label{e4}
\end{eqnarray}
One can demonstrate that (\ref{e2}) and (\ref{e4}) are consistent.
To each physical quantity $A$
 given in the Hamiltonian formalism by the
 function
$A(\eta,t)$,
corresponds
a ``quasi--Schr\"odinger'' operator  $\tilde{A}$ by the rule
$\tilde{A} = A(\tilde{\eta},t)$;
in the same manner one constructs the quantum
Hamiltonian  $\tilde{H}$, according to the classical one $ H(\eta,t)$.
The time evolution of the state vectors  $\Psi$ in this  picture
is determined by the Schr\"odinger
equation with the Hamiltonian $\tilde{H}=
H(\tilde{\eta},t)$.  The total time evolution results both from the evolution
of the state vectors and from one of the operators. It is convenient to analyse
such an evolution in the Heisenberg picture whose operators $\check{\eta}$
are related to the operators $\tilde{\eta}$ as $\check{\eta}=U^{-1}
\tilde{\eta}U$, where $U$ is the evolution operator related to the Hamiltonian
$\check{H}$. Such operators satisfy the equations
\begin{eqnarray}\label{II.18a}
&&\dot{\check{\eta}}=\left.\{\eta,H +
\epsilon\}_{D(\phi)}\right|_{\eta=\check{\eta}},\\
&&[\check{\eta},\check{\eta}']=i\left. \{\eta,\eta'\}_{D(\phi)}\right|_
{\eta=\check{\eta}}~, \hskip 5mm
\phi(\check{\eta},t)=0  \; . \nonumber
\end{eqnarray}
All the relations (\ref{II.18a}) together may be considered as a prescription
for quantization in the Heisenberg picture for theories with non-stationary
second-class constraints. The total time evolution is controled only by the
first set of the equations (\ref{II.18a}) since the state vectors do not depend
on time in the Heisenberg picture. In the general case such an evolution is not
unitary. Suppose, however, that a part of the set of second-class constraints
consists of supplementary gauge conditions, the choice of which is in our
hands. In this case one may try to select these gauge conditions in a special
form to obtain unitary evolution. The evolution is unitary if there exists
an effective Hamiltonian $H_{eff}(\eta)$ in the initial phase space of the
variables $\eta$ so that the right side of the equations of motion (\ref{e1})
may be written as follows
\begin{equation}
\dot{\eta}=\{\eta,H + \epsilon\}_{D(\phi)}=\{\eta,H_{eff}\}_{D(\phi)}\;.
\label{II.18b}
\end{equation}
In this case, (due to the commutation relations (\ref{II.18a})) the quantum
operators $\check{\y}$ obey the
equations (we disregard here problems connected
with operator ordering)
\be\label{II.20}
\dot{\check{\y}} = - i [ \check{\y}, \check{H}_{eff} ], \quad
\check{H}_{eff}=H_{eff}(\check{\y})~.
\ee
The latter allows one to introduce the real Schr\"odinger picture
where operators do not depend on time but the evolution
is controlled by the Schr\"odinger equation with the Hamiltonian $H_{eff}$.
We may call the gauge conditions which imply the existence of the effective
Hamiltonians as {\it unitary gauges}.
Remember that in the stationary constraint
case all gauge conditions are
unitary \cite{g-t}.
As it is known \cite{g-t}, the set of second-class constraints
can always be
solved explicitly with respect to part of the variables
$
\y_{*} = \Psi(\y^*), \hskip 2mm \y = (\y_*, \y^*), 
$
so that $\y_*$ and $\y^*$ are sets of pairs of
canonically conjugated variables $ \y_{*} = (q_{*}, p_{*}),\hskip 2mm
\y^{*} = (q^{*}, p^{*}).$
We may call $\y^*$ as independent variables and $\y_*$ as
dependent ones.
In fact $\y_* - \Psi(\y^*) =0 $ is an equivalent to
$\f(\y) =0 $ set of second-class constraints.
One can easily demonstrate that it is enough to verify the existence
of the effective Hamiltonian (the validity of relation (\ref{II.20}))
for the independent variables only.
Then the evolution of the dependent variables which is controlled
by the constraint equations is also unitary.

In the situation of our main interest here, when the Hamiltonian is
proportional to the constraints, one can put $H=0$ in the equations
(\ref{II.18a}). Thus, the ``quasi--Schr\"odinger'' picture and the Heisenberg
one coincide. The time evolution is unitary in this case if the following
equations hold
\begin{eqnarray}
\dot{\eta}=\{\eta,\epsilon\}_{D(\phi)}=-\{\eta,\phi_{a}\} C_{ab}
\frac{\partial\phi_{b}}{\partial t}=\{\eta,H_{eff}(\eta)\}_{D(\phi)}.
\label{II.19}
\end{eqnarray}

Let us analyse the theory (2.9) in the gauge (\ref{m16}) using the
above consideration.  The matrix $\{ \f_a, \f_b \}$ is simple in this case:
$\{ \f_a, \f_b \} = {\rm antidiag} (-1,1), ~ C_{ab} =\{ \f_b,
\f_a \}$.  The Dirac brackets between the independent variables
${\bf x},{\bfp}$
are reduced to the Poisson ones,
\be
\{ x^i, x^j \}_D = \{ \p_i, \p_j \}_D =
0~, \hskip 5mm \{ x^i, \p_j \}_D = \d^i_j~.
\ee
The time evolution of these
variables is given by the equations
\be
\dot{{\bf x}} = - \{ {\bf x}, \f_a \} C_{ab}
\dot{\f}_b = \{{\bf x}, H \}~, \hskip 5mm \dot{{\bfp}} = - \{{\bfp},
\f_a \} C_{ab} \dot{\f}_b = \{ {\bfp}, H \}~,
\ee
where $H$ is the
Hamiltonian of the theory (\ref{m1}) and at the same time it is the effective
Hamiltonian in our definition.  This means that in the
chronological gauge the dynamics of the original non-singular  theory is
reproduced.

Let us consider instead of (\ref{m16})
a more general gauge fixing  $\f_2 = x^0 - \v({\bf x}, \bfp, t) =0$.
To get conditions   on the function $\v$, which
make the gauge unitary, we restrict ourselves
to the free particle case, where $H$ from (\ref{m10}) is
${\bf p^2}/2m$.
In this case $\{ \f_a, \f_b \} = {\rm antidiag} (-K,K), ~ C_{ab} =
K^{-2} \{ \f_b, \f_a \},~
K = \left( 1 - \frac{\p_i}{m} \part_i \v \right)$. The non-zero Dirac
brackets between the independent variables ${\bf x},{\bfp}$ are:
\be
\{ x^i, x^j \}_D = (mK)^{-1}
\left( \frac{\part \v}{\part \p_i} \p_j -
\frac{\part \v}{\part \p_j}  \p_i \right),              \label{n9}
\hskip 4mm
\{ x^i, \p_j \}_D =  \d^i_j + (mK)^{-1} \p_i \part_j \v. 
\ee
According to (\ref{II.19})
these variables obey the following equations:
\be
\dot{{\bf x}} = - \{{\bf x}, \f_a \} C_{ab} \dot{\f}_b =
(mK)^{-1}{\bf\p}\dot{\v}~,                             \label{n12}
\hskip 5mm
\dot{{\bfp}} =  - \{{\bfp}, \f_a \} C_{ab} \dot{\f}_b =0~. 
\ee
On the other hand, if the effective Hamiltonian $H_{eff}$
does exist (unitary gauge), one can write
\begin{eqnarray}
\dot{x}^i &=& \{ x^i, H_{eff}\}_D =  (mK)^{-1}
\left( \frac{\part \v}{\part \p_i} \p_j -
\frac{\part \v}{\part \p_j}  \p_i \right) \part_j H_{eff} +
\left[ \d^i_j + (mK)^{-1} \p_i \part_j \v \right]
\frac{\part H_{eff}}{\part \p_j}\;,            \nonumber 
\\
\dot{\p_i} &=& \{ \p_i, H_{eff}\}_D = -
\left( \d^i_j + (mK)^{-1} \p_j \part_i \v \right) \part_j H_{eff}\;.
                                                          \label{n15}
\end{eqnarray}
Comparing (\ref{n12})  with  (\ref{n15})
we get the following conditions on $H_{eff}$:
\be
\part_j H_{eff}=0, \hskip 5mm \frac{\part H_{eff}}{\part \p_i}
= \p_i  (mK)^{-1} \left( \dot{\v} -
\frac{\part H_{eff}}{\part \p_i} \part_i\v \right)~.       \label{n17}
\ee
The first eq. (\ref{n17}) means that $H_{eff}$ does not depend
on ${\bf x}$ and the second one results in the condition
\be
\left(\p_j \frac{\part }{\part \p_i} -
\p_i \frac{\part }{\part \p_j}   \right) H_{eff} =0~,
\ee
which means that $H_{eff}$ depends only on ${\bfp}^2$.
Thus, $H_{eff} = H_{eff}({\bfp}^2, t)$.
Using this information in the second equation (\ref{n17}),
we get:
\be
2m \frac{\part H_{eff}}{\part \bfp^2} = \dot{\v}~.
\ee
Thus, $\dot{\v}$ is a function on $\bfp^2$ and $t$ only.
That leads to  the following structure:
\be
\v({\bf x}, {\bfp}, t) = \chi({\bf x},\bfp) + \psi(\bfp^2,t)~,
\label{n19}
\ee
where $\chi$ and $\psi$ are arbitrary functions on the
indicated arguments.
The effective Hamiltonian in this case can be expressed via the
function $\psi(\bfp^2,t)$ only:
\be
H_{eff} = \frac{1}{2m} \int \dot{\psi}({\bfp}^2,t)
 d{\bfp}^2~.          \label{n20}
\ee

As an example of a nonlinear in time $t$ gauge condition we consider
here the following
\be
\f_2 = x^0 - t  - \frac{ma}{2\pi} t^2  = 0, \label{nonlin}
\ee
where for simplicity we have selected one-dimensional case, i.e. the
Hamiltonian of the initial nonsingular theory is $H = \pi^2/2m $.
The previous consideration is valid in this case, thus, (\ref{nonlin})
is an unitary gauge which generates  the
effective Hamiltonian of the form
\be
H_{eff}= \frac{\pi^2}{2m} + \pi at~.
\ee
If we suppose that the initial non-singular action (\ref{m1})
corresponds to
a theory in an inertial reference frame, then the chronological gauge
(\ref{m16}) returns us to the description in such a frame, whereas the gauge
(\ref{nonlin}) corresponds to the description from the point of view
of an accelerating (with acceleration $a$) frame.

Let us turn to the question about physical quantities in the RI theory
under consideration. It is known \cite{ab1,g-t} that in conventional
gauge theories physical quantities, which are defined by functions on
the phase space, have to commute with first-class constraints on the
mass shell (Dirac's criterion). What kind of restrictions does  this criterion
impose on the physical quantities in our case? Due to the constraint
(\ref{m13}), the physical quantities, which are given by functions on the
phase space of variables $x^{\m}, \p_{\m}$, always may be expressed via
functions of the form $A = A(x^0, \bfy),~~ \bfy = ({\bf x}, \bfp)$.  The
condition of commutativity of such functions with the first-class constraint
(\ref{m13}) on the mass shell results then in
\be
\{ A, \f_1 \} = \frac{\part
A}{\part x^0} + \frac{\part A}{\part \bfy} \{ \bfy, H\} \approx 0~.
\label{star}
\ee
Remembering, that the equations of motion in the theory
under consideration have the form
\be
\dot{ \bfy} = \{ \bfy, H^{(1)} \} = \l \{ \bfy, H \}~,
\hskip 5mm
\dot{x}^0 = \{ x^0, H^{(1)} \} = \l~,
\ee
we may rewrite (\ref{star}) as
\be
\frac{\part A}{\part x^0} \dot{x}^0 + \frac{\part A}{\part \bfy}
\dot{\bfy} = \frac{d A}{dt} \approx 0~. \label{starstar}
\ee
Thus, the Dirac's criterion admits as physical functions only those which
present integrals of motion. We believe that the RI theory under
consideration in the chronogical gauge (\ref{m16}) has to coincide with the
initial non--singular theory (\ref{m1}), in which all the functions of the
form $A=A(t,{\bf\eta})$ are physical. Thus, if one accepts the Dirac's
criteria then an essential part of real physical quantities of the initial
non--singular theory (\ref{m1}) are lost and the RI version is not
equivalent to the initial theory.

The above consideration looks even more transparent in
the case of the field theory.
Let us consider, for example, a theory of a scalar field in a flat
space-time.
The action of the theory being written in an inertial RF has the form:
\be
S  = \int {\cal L} d^{D+1} x =
\int  \left[ \frac12 \y^{\m\n} \v,_{\m} \v,_{\n} + F(\v) \right]
d^{D+1} x~,                   \label{1}
\ee
where $\y_{\m\n} = {\rm diag}( 1,-1,\ldots,-1)$,
$F(\v)$ are some terms independent of the derivatives
of $\v$, and $\v,_{\m} = \part \v/\part x^{\m}$.
Let us change in (\ref{1}) $x^{\m}$ to $y^{\m}$ and then let us
rewrite the integral   in the RHS (\ref{1}) doing the substitution
$y^{\m} = y^{\m}(x)$. Thus, we get
\be
S_R = \int {\cal L}_R d^{D+1} x
 = \int  \left[ \frac{1}{2} g^{\m\n}  \v,_{\m}  \v,_{\n}
+ F(\v) \right] \sqrt{-g} d^{D+1} x~,                     \label{action.rep}
\ee
where
\be
g^{\m\n} = a^{\m}_{\a} a^{\n}_{\b} \y^{\a\b}~, \hskip 5mm
a^{\m}_{\a} y,_{\n}^{\a} = \d_{\n}^{\m}~, \hskip 5mm
g = det \mid \mid g_{\m\n} \mid \mid = - e^2~, \hskip 5mm
e =   det \mid \mid y,_{\n}^{\m} \mid \mid~,       \label{metric}
\ee
and $g_{\m\n}$ is the inverse of $g^{\m\n}$.
If one treats the $y^{\m}$ as four  new scalar fields,
then the theory becomes a gauge one, with the corresponding
gauge transformations having the form:
\be
\d y^{\m} =   y,_{\a}^{\m} \d \z^{\a}~,
\hskip 5mm \d \v  =  \part_{\a} \v \d \z^{\a} ~,
\ee
where $\d \z(x)$ are $D+1\;x$-dependent parameters of
the gauge transformations.
To see the relation between the theories (\ref{1}) and
(\ref{action.rep}) we construct their Hamiltonian versions
as in the previous finite-dimensional case. Let us start with the gauge
theory (\ref{action.rep}). Using the relations
\be
\frac{\part e}{\part \dot{y}^{\a}} = ea_{\m}^0~, \hskip 5mm
\frac{\part g^{\m\n}}{\part \dot{y}^{\a}} = -2 g^{0\m}
a_{\a}^{\n}~,
\ee
we introduce the canonical  momenta:
\[ \p = \frac{ \part
{\cal L}}{\part \dot{\v}} = e a^0_{\m} a^0_{\n} g^{\m\n} \dot{\v} + e
a^0_{\m} a^i_{\n} g^{\m\n} \v,_i~,  \hskip 5mm \p_{\m} = \frac{\part {\cal
L}}{\part \dot{y}^{\m}} = - \frac{1}{2} e a^0_{\m} a^0_{\n} a^0_{\r} g^{\n\r}
\dot{\v}^2 - \] \be -e a^i_{\m} a^0_{\n} a^0_{\r} g^{\n\r} \dot{\v} \v,_i + e
 \left[ \frac{1}{2}  a^0_{\m} a^i_{\n} a^j_{\r} - a^i_{\m} a^j_{\n} a^0_{\r}
\right] g^{\n\r}  \v,_i  \v,_j + e a^0_{\m} F(\v)~.\label{momen}
\ee
Equations (\ref{momen}) allow one to express only the
velocity $\dot{\v}$ via fields and momenta,
velocities $\dot{y}^{\m}$ remain unexpressible,
\be
\dot\v = \frac{\p - e  a^0_{\m} a^i_{\n} g^{\m\n}
  \v,_i}
{e a^0_{\m} a^0_{\n} g^{\m\n}}~.
\ee
Thus,  the primary constraint $\f_1 =0$ appear:
\be
\f_{1\m} = \p_{\m} + a_{\m}^{\a}(y) {\cal H}_{\a}(y)~,  \label{constr2}
\ee
where
\begin{eqnarray}
{\cal H}_0(y) &=& \frac{\p^2}{2 e g^{00}} - \frac{g^{oi}}{g^{00}} \v,_i \p
- \frac{e}{2} \frac{\g^{ij}}{g^{00}} \v,_i \v,_j -e F(\v)~, \\
{\cal H}_i(y) &=& \v,_i \p~, \hskip 10mm
\g^{ij} = - \frac{g^{0i} g^{0j}}{g^{00}}+g^{ij}~.
\end{eqnarray}
The density of the total Hamiltonian is
\be
{\cal H}^{(1)} = \l^{\m} \f_{1\m}~, \hskip 5mm
\l^{\m} = \dot{y}^{\m}~,
\ee
where the unexpressible velocities $\dot{y}^{\m}$
appear as Lagrange multipliers. No more constraints appear and $\f_1$ are
the first--class constraints. A possible form of the gauge conditions is
\be
\f^{\m}_2 = y^{\m} - f^{\m}(x) =0~, \hskip 5mm
\left| \frac{\part f}{\part x} \right| \neq 0~.
          \label{fix.gen}
\ee
Together with the primary constraints they form a
set of second-class constraints,
which can be written in the following equivalent
form $\F = 0$, where
\be
\F =  \left\{ \begin{array}{ll}
\p_{\m} + a_{\m}^{\a}(f(x)) {\cal H}_{\a}(f(x)) =0 ~, \\
y^{\m} - f^{\m}(x)= 0~.
\end{array}
\right.                  \label{third.lorentz}
\ee
One can select $Q = (\v, \p)$  as independent variables.
The Dirac brackets between them are:
\be
\{ \v, \p \}_{D(\F)} = \{ \v, \p \} = 1~,
\hskip 5mm
\{ \v, \v \}_{D(\F)} = \{ \v, \v \} = 0~,
\hskip 5mm
\{ \p, \p \}_{D(\v)} = \{ \p, \p \} = 0~. \label{dirac.bra}
\ee
The time evolution is given by an effective Hamiltonian,
\begin{eqnarray}
&&\dot{Q} = - \{ Q, \F_A\} C_{AB} \dot{\F}_B
= \{ Q, H_{eff} \}~, \hskip 5mm C_{AB} = \{ \F, \F \}^{-1}_{AB}~,
\nonumber \\
&& H_{eff} =
 \int {\cal H}_0(f(x)) {\bf dx} ~.
\label{time3}
\end{eqnarray}
Thus, the gauge (\ref{fix.gen}) is unitary.
One can easily see that the equations of motion
(\ref{time3}) reproduce the dynamics of the initial
theory of scalar field in flat space,
but in a curvilinear RF, the coordinates $x$ of which
being related to the coordinates $y$ of the inertial
RF by the transformation (\ref{fix.gen}).
If $f^{\m}(x) = x^{\m}$ (an analog of the
chronological gauge (\ref{m16}) of the finite-dimensional case),
or $f^{\m}(x) = \L^{\m}_{\n} x^{\n},~ (\L^T \y \L = \y)$,
then we get back to the initial theory in an inertial  RF.
In this case the effective Hamiltonian (\ref{time3})
takes the familiar form:
\be
H_{eff} = \int {\cal H} {\bf dx} = \int \left[ \frac12 (\p^2 +
\v,_i^2)  - F(\v)\right] {\bf d x}~.     \label{hamilt}
\ee
What are physical quantities in the theory (\ref{action.rep})?
The Dirac's criterion admits only those ones which commute with all
first--class constraints.  In our case, that would mean:
\be
\{ A, \f_{1\mu} \}
\approx 0~,         \label{condition}
\ee
where $\f_1$ is given in
(\ref{constr2}).  Due to the same constraint (\ref{constr2}) the physical
quantities, which are functions on the phase space,
may always be taken in the form $A = A(y,{\bfy}),\;{\bfy}=(\v, \p)$.
For such functions the condition (\ref{condition}) results in:
\be
\{ A, \f_{1\m} \} = \frac{\part A}{\part y^{\m}}
+ \frac{\part A}{\part \bfy} a^{\a}_{\m} \{ \bfy, H_{\a} \}
\approx 0~.
\ee
Multiplying this equation by the nonsingular matrix $y,_{\b}^{\m}$
one obtains the following relation:
\be
\frac{d A}{d x^{\m}} \approx 0~,         \label{fiz.4dim}
\ee
which is the generalization of the finite-dimensional
equation (\ref{starstar}).
Equation (\ref{fiz.4dim}) means that the
above criterion admits as physical only functions that do not
depend on space-time.

Similar to the finite--dimensional case we meet here the following situation.
If we accept the Dirac's criterion then we can not identify the RI version of
the scalar field theory with the initial formulation in flat space time even
in the ``chronological'' gauge. That circumstance indicates us that the above
criterion has to be critically reconsidered in the situation under
consideration (see for detailed discussion the next section).

\section{Relativistic particle theory. RI and time inversion}

In this Section we are going to discuss theory of a relativistic
particle as an instructive example of RI system. Such a theory is
interesting by itself and has attracted attention already for a long time,
in particular, due to the fact that it can serve as a prototype for a
string theory (now one can consider it as 0-brane theory). On this example we
are going to study different possibilities of time dependent gauge fixing and
a relation between reparametrizations and time-inversion symmetry.

Let us restrict ourselves for simplicity to spinless
particles moving in an external electromagnetic field
with the potentials $A^{\m} = (0, {\bf A}({\bf x}))$,
which corresponds to the case of a constant magnetic field.
The theory of such a particle is described by the action
\cite{LL}:
\be
S = \int \left[ -m \sqrt{1 - (\dot{\bf x}^2)} + g \dot{\bf x}{\bf A}
\right] dt~,
                                                               \label{r1}
\ee
where ${\bf x} = (x^i)$ are spatial coordinates of some inertial
reference frame and $t$ is the time of the same frame, $g$ is the algebraic
charge of the particle and $m$ its mass.
The action (\ref{r1}) is non--singular, so that hamiltonianization and
quantization can be done directly.
The three--dimensional momentum vector $\bfp$
is defined by the relation:
\be
\bfp = \frac{\part L}{\part \dot{\bf x}} =
\frac{m \dot{\bf x}}{\sqrt{1 - (\dot{\bf x}^2)}} + g {\bf A}~,
\hskip 5mm {\bfp} = (\p_i)~.         \label{r2}
\ee
The classical equations of motion are:
\be
\dot{\bfy} = \{ \bfy, \o \}~, \hskip 5mm
\bfy = ({\bf x}, \bfp)~,  
\hskip 5mm  \o = \sqrt{m^2 + ( \bfp - g {\bf A})^2}~.\label{r6}
\ee
They describe the motion of a  particle with
charge $g$ in the constant magnetic field.
Going over to the quantum theory we get the commutation
relations between the operators $\hat{\bf x}, \hat{\bfp}$:
$
[\hat{x}^i, \hat{\p}_k] = i  \{ x^i, \p_k \} = i \d^i_k$.
In the coordinate representation $\hat{\bf x}$ is a
multiplication operator, whereas
$ \hat{\bfp} = - i \frac{\part}{\part {\bf x}}$.
The state vectors $\psi$ obey the Schr\"odinger
equation
\be
i \frac{\part \psi}{\part t} = \hat{\o} \psi,
\hskip 5mm
\hat{\o} = \sqrt{m^2 +(i {\bf \nabla}+ g {\bf A})^2} ~.
                                                  \label{r8}
\ee
The quantum theory constructed in this way describes only one particle
with charge $g$. Such a theory is not equivalent to the theory
which is based on the Klein-Gordon
equation.
Indeed, the latter describes states of  charged particles
with positive and negative energies or states of particles
and antiparticles (charge $(-g)$) with positive energies.

Let us consider a RI formulation of the system in question. The corresponding
action has the form
\be
S = \int [ -m \sqrt{\dot{x}^2} -
g \dot{x}^{\m} A_{\m}] d\t~, \hskip 5mm
\dot{x}^{\m} = \frac{ d {x}^{\m}}{d \t}~,         \label{r9}
\ee
where now four $x^{\mu}=(x^0,{\bf x})$ are dynamical variables
dependent on a new time $\tau$.
 The action (\ref{r9}) similar to the one (\ref{action1})
 obeys the reparametrization gauge symmetry (\ref{II2}).
Hamiltonianization and quantization of the theory is more complicated
than in the previous case.
Let $\p_{\m}$ be the generalized momenta related to the
variables $x^{\m}$,
\be
\p_{\m} = \frac{\part L}{\part \dot{x}^{\m}} =
- \frac{m \dot{x}_{\m}}{\sqrt{\dot{x}^2}} - g A_{\m}~.       \label{r10}
\ee
Then there is a constraint $(\p + g A)^2 = m^2$,
which can be written in the following equivalent
form, which is convenient for our purposes:
\be
\f_1 = \p_0 + \z \o = 0, \hskip 5mm
\z = - {\mbox{sign}}~ \p_0~.
\label{a}
\ee
One can express from (\ref{r10}) three velocities $\dot{\bf x}$
as well as the sign of $\dot{x}^0$ in terms of the coordinates,
momenta, 
and one unexpressible velocity, which  is here
$\l = \mid \dot{x}^0 \mid$,
\be
\dot{\bf x} = \l \o^{-1} (\bfp - g {\bf A})~,
\hskip 5mm
{\mbox {sign}}~ \dot{x}^0 = \z~,
\hskip 5mm
\sqrt{\dot{x}^2} = m \l \o^{-1}~.                           \label{b}
\ee
Thus, one can construct the total
Hamiltonian $H^{(1)}$ by substituting (\ref{b}) in the expression
$\p_{\m} \dot{x}^{\m} - L $,
\be
H^{(1)} = \l \z \f_1  ,                             \label{r16}
\ee
where $\l$  is a Lagrange
multiplier subjected, however, to the condition of positivity.
The Hamiltonian equations of motion of the form
\be
\dot{x}^{\m} = \{ x^{\m}, H^{(1)} \}, \hskip 5mm
\dot{\p}_{\m} = \{ \p_{\m}, H^{(1)} \}, \hskip 5mm
\f_1 = 0,  \hskip 5mm \l \geq 0~,                      \label{r17}
\ee
are equivalent to the Lagrangian ones. 
No secondary constraints arise from the consistency conditions
and $\l$ remains undetermined.
This indicates that we are dealing with a gauge theory.
The total Hamiltonian is proportional to the constraints,
as one can expect for a RI theory.
Below we are going to
discuss some possible  gauges and quantization in these
gauges.

First, let us consider the case of a neutral ($g=0$) particles. In this
case the action (\ref{r9}) is invariant under the time inversion $\tau
\rightarrow -\tau $. Since the gauge symmetry in the case under consideration
is related to the invariance of the action under the changes of the variables
$\tau$, there appears two possibilities: namely, to include or not to include
the above discrete symmetry in the gauge group together with continuous
reparametrizations. Let us first study the former possibility and include the
time inversion in the gauge group. Then the gauge conditions have to fix the
gauge freedom which corresponds to both kind of symmetries, namely, to fix
the variable $\lambda=|\dot{x}^0|$, which is related to the
reparametrizations, and to fix the variable $\zeta={\rm
  sign}\,\dot{x}^0$, which is related to the time inversion. To this
end we may select the chronological gauge of the form
\be
\f_2 = x^0-\t =0\;.                              \label{nonrel.chron}
\ee
The consistency condition $\dot{\f}_2 =0$ leads
on the constraint surface to the equation
\be
\dot{\f}_2 =  \frac{\part \f_2}{\part\t} +
\{\f_2, H^{(1)} \} = -1 + \l \z  = 0~,
\ee
which results in the condition $\z \l =1$.
Remembering that $\l \geq 0$, we get $\z =1,\lambda=1$.
That reduces the constraint surface to the following form:
$\f_a = 0,~a=1,2,$
\be
\f_1 = \p_0 + \o~, \hskip 5mm
\f_2 =x^0 - \t~.
\ee
It is easy to calculate that $\{ \f_a, \f_b \} =
{\rm antidiag}~(-1, 1)$ and $C_{ab} = - \{ \f_a, \f_b \},~
C_{ab} \{ \f_b, \f_c \} = \d_{ac}$.
One can select $\bfy = ({\bf x}, \bfp)$ as independent
 variables. 
Their Dirac brackets coincide with the Poisson ones, 
\be
\{ \bfy, \bfy'  \}_D =\{ \bfy, \bfy'  \}~.
\label{commutD}
\ee
The quantum operators $\check{\bfy}$ obey the equation (\ref{II.18a}),
which in this particular case takes the following form
\begin{eqnarray}\label{ido}
&&\dot{\check{\bfy}} = - \{ \bfy, \f_a \} C_{ab}
\left.\frac{\part \f_b}{\part \t}\right|_{\bfy = \check{\bfy}}
= \left.\{ \bfy, \o \}\right|_{\bfy = \check{\bfy}} =
-i  [ \check{\y}, \check{\o} ]~,\\  
&&[\check{\bfy},\check{\bfy}']=i\{\bfy,\bfy'\}.   \nonumber
\end{eqnarray}
Thus, the evolution is unitary and is governed by the effective
Hamiltonian $\o$ (\ref{r6}).
One can consider time independent Schr\"odinger operators
$\hat{\bfy} = e^{-i\check{\o}\t} \check{\bfy}(\t) e^{i\check{\o}\t} $
and time dependent state vectors. 
The operators $\hat{\bfy}$ obey the same commutation relations
(\ref{ido}) and can be realized as in the non-reparametrization
invariant case.
Thus, we get the Schr\"odinger equation (\ref{r8}) if one identifies
$\t$ with $t$.

Suppose we do not include the time inversion in the gauge group. That is
espesially natural when $g\neq0$, $A_\mu\neq0$, because in this case the time
inversion is not anymore a symmetry of the action. Thus, one may now
consider more general situation of the charged particle moving in the
external magnetic field. Under the above supposition the condition
(\ref{nonrel.chron}) is not anymore a gauge,
it fixes not only the reparametrization gauge freedom (fixes
$\lambda $) but it fixes also the variable $\zeta$ which is now
physical. A possible gauge condition has the form \cite{ab3}:
\be
\f_2 = x^0 - \z\t =0~.                         \label{relat.chron}
\ee
The consistency condition $\dot{\f}_2 = 0$ leads to the equation
\be
\dot{\f}_2 = \frac{\part \f_2}{\part \t} + \{ \f_2, H^{(1)} \}
 = -\z +\l \z =0~, 
\ee
which fixes only $\l =1$ and retains $\z$ as a physical variable.
Trajectories with $\z = +1$ correspond to particles, while
 trajectories
with $\z = -1$ to antiparticles \cite{ab3}.
Two second-class constraints
\be
\f_1 = \p_0 + \z \o\;, \hskip 5mm
\f_2 = x^0 - \z \t\;,
\ee
form the same algebra like in the previous case.
One has only to add the relation $\{ \z, \bfy \}_D = 0$
to the Dirac brackets (\ref{commutD}).
However, we get here an additional operator $\hat{\z}$,
which has to be realized in the Hilbert space of state vectors.
We assume the operator $\hat{\z}$ to have the eigenvalues
$\z = \pm 1$ by analogy with the classical theory.
Such an operator can be realized in a Hilbert space whose elements
are two-component columns
\be
\Psi = \left(\begin{array}{c}
\Psi_1({\bf x}) \\
\Psi_2({\bf x})
\end{array}
\right),
\ee
if we chose the operator  $\hat{\z}$ as the matrix
$\hat{\z} = {\rm diag} (1,-1)$.
The time independent operators $\hat{\y}$ can be realized as follows
\be
\hat{x}^i = x^i {\bf I}~, \hskip 5mm
\hat{\p}_j = - i \part_j {\bf I}~,
\ee
where ${\bf I}$ is a unit $2 \times 2$ matrix.
The time evolution of the state vectors is described by the
Schr\"odinger equation
\be\label{zz}
i \frac{\part \Psi}{\part \t} = \hat{\o} \Psi,
\ee
where $\hat{\o}$ is given by eq.(\ref{r8}).
The equation (\ref{zz}) differs from the similar equation (\ref{r8}) due to
the structure of the Hilbert space, which now allows one to describe states
for both particles and antiparticles.

As an example of gauge conditions which lead to the description from the
point of view some non--inertial reference frames we consider here the gauge
($a=$const)
\be\label{IV.22}
\f_2 = x^0 +{\pi_0\over m}\t+a=0
\ee
in case when the time inversion is not included in the gauge group and the
gauge
\be\label{IV.22a}
\f_2 = x^0 -{|\pi_0|\over m}\t+a=0
\ee
when it does.

One can demonstrate first that the gauge condition
(\ref{IV.22}) corresponds (at any $a$) to the proper-time gauge $\dot{x}^2
=1$ in the Lagrangian formulation.  Indeed the consistency condition \be
\dot{\f}_2 = \frac{\part \f_2}{\part \t} + \{ \f_2, H^{(1)} \}
= \frac{\p_0}{m} + \l \z = 0~,             \label{26.2}
\ee
defines $\l = \frac{ \mid \p_0 \mid}{m}$.
Remembering the last relation (\ref{b})
and the constraint (\ref{a}) we can see that (\ref{IV.22}) at any $a$ is
equivalent to the condition $\dot{x}^2 =1$. Thus, (\ref{IV.22}) may be called
proper--time gauge in Hamiltonian formulation.
The proper--time gauge,  similarly to the chronological
gauge (\ref{relat.chron}),
does not fix the variable $\z$, and leaves the possibility
to describe particles and antiparticles at the same time.
The gauge condition (\ref{IV.22a}) 
similar to one (\ref{nonrel.chron}) fixes the
variables $\zeta$, thus it is acceptable only when the time inversion
(\ref{IV.22})is
included in the gauge group.

The constraint algebra in both gauges (\ref{relat.chron}) and (\ref{IV.22}) is
the same, the commutation relations and the realization for the independent
operators are also the same, however the effective Hamiltonian in the
proper--time gauge is different,
\be
H_{eff} = \frac{\o^2}{2m}~.\label{26.3}
\ee
Thus, the Schr\"odinger equation has the form
\be
i\frac{\part \Psi}{\part \t} = \frac{ \hat{\o}^2}{2m} \Psi\;.
\ee
One can establish a formal relation between the gauges (\ref{relat.chron}) and
(\ref{IV.22}). Namely, one can present a canonical transformation, which
connects both gauges on the classical level.  The generating function of a
such transformation has the form:
\be
W = x^{\m}\p'_{\m} + \t \mid \p'_0 \mid -\t \frac{{\p'_0}^2}{2m}~, \label{26.5}
\ee
if the phase space variables without the primes are related to the
chronological gauge (\ref{relat.chron}) and the primed ones to the
proper--time gauge (\ref{IV.22}).  The transformation does not change the
variables $x^i$ and $\p_{\m}$.  It changes only
$x^0,\hskip 2mm x'^0 = x^0 - \z \t - \frac{\p_0}{m}\t$.  Thus, it transforms
the constraint surface of the first
gauge into the one of the second gauge.  One can also see that this
transformation connects both Hamiltonians
\be
H = H' + \frac{\part W}{\part \t} = \frac{\p'^2_0}{2m} + \mid \p_0'\mid -
\frac{\p'^2_0}{2m} = \mid \p'_0 \mid = \mid \p_0 \mid = \o~.
\ee
On the quantum level the state vectors in
both gauges are connected by means of a quantum canonical transformation \be
\Psi = e^{ - i \hat{W}} \Psi'~, \hskip 5mm
\hat{W} = \t \hat{\o} - \t \frac{\hat{\o}^2}{2m}~.
\ee

In the spirit of the interpretation given in Sec.3 we may say that the
chronological gauges (\ref{nonrel.chron}) and (\ref{relat.chron}) lead us to
the inertial RF, whereas the proper--time gauges (\ref{IV.22})
and (\ref{IV.22a}) correspond to the description from the point of view of
non--inertial (at $A\neq0$) RF. A formal possibility to connect
these two gauges by means of a canonical transformation does not mean their
physical equivalence since such a transformation depends explicitly on time.

\setcounter{equation}{0}
\section{Possible interpretation}
Results of the consideration which was presented in two previous
Sections may be summarized in the following generalizing
interpretation. Let us turn first to the non-RI actions (2.7),
(2.38), and (3.1).It is natural to believe that such  actions
give descriptions of the corresponding physical systems in
certain RF.
For example, actions (2.38) and (3.1) provide a description from
the point of view of an inertial RF with a Cartesian base.
Constructing RI versions of the above mentioned actions we see
that a possibility appears to describe the same physical system
from the point of view of a more wide class of RF. The theories
become gauge ones, they contain additional non-physical
variables. The corresponding gauge symmetry - RI leads always to
zero Hamiltonian phenomenon. To introduce a dynamics we fix a
gauge by means of supplementary conditions which depend on time
(or space-time variables) explicitly. It turns out that such a
gauge fixing looks literally like a certain choice of a RF. In
particular, the chronological gauges correspond to the RF in
which initial non-RI actions are formulated. More complicated
gauges reproduce in general non-inertial curvilinear RF. Based
on the experience that was derived from the simple example
consideration we believe that any fixation of the
reparametrization gauge freedom corresponds always to a
certain choice of the space-time RF. Here we have especially
emphasized the origin of the RF which is fixed. The point is
that the fixation of the gauge freedom of any kind may be
trreated as a choice of some RF. In this sense the
reparametrization symmetry is similar to gauge symmetries of
different nature, let us call them internal gauge symmetries
(one may define the latter symmetries as ones which do not
involve the space-time coordinate transformations). The
principle distinction between the reparametrizations and
internal gauge symmetries are related with the distinction
between the corresponding RF. Whereas one believes that the RF
for the internal gauge symmetries may not be realized physically
(at least until now), the choice of RF to measure space-time
coordinates may be physically realized. If in the former case
the physical quantities do not depend on the choice of the
gauge, in the latter case this may be not true. To describe local
physical quantities it is natural to use space-time dependent
functions which depend explicitly on the choice of RF and are
transformed in a certain way under the RF change. Thus, we have
to admit gauge non-invariant objects to decribe physics. As it
is known \cite{ab1,g-t}, when the gauge transformations do not involve a
 transformation of space-time coordinates, gauge invariant functions on the
 phase space have to commute with first-class constraints on the mass shell
 (Dirac's criterion).  The previous reasonings mean that the "local" point of
view, which is, in fact, advocated here, abrogate the Dirrac's criterion with
respect to the first-class consraints which generate the reparametrizations.
Rejection of the Dirac's criterion in the case of the reparametrization gauge
symmetry admits, thus, any functions (which are physical with respect to the
internal gauge symmetries) as physical ones. The choice of them is dictated
by  concrete conditions of the problem. Let us, for example, return to the
theory of scalar field studied in Sect.II. Let us have a Lorentz tensor in
the initial non-RI formulation, let say the vector $\varphi_{,\mu}(x)$. The
question is: what kind of physical quantity corresponds to it in
the RI formulation? One may present two naturally constructed
quantities, the general coordinate vector $\varphi_{,\mu}(x)$
and the scalar $a^{\mu}_{\alpha}\varphi_{,\mu}(x)$. Both of them
coincide with the initial physical quantity in the chronological
gauge (in the inertial RF). 
In the literature one may often meet some arguments
in favor of the latter choice
 (see for
example \cite{feng}).

We know that gauges which fix an internal gauge symmetry may
always be selected in time (space-time) independent form
(canonical gauges). Such gauges then may be related by means of
a time-independent canonical transformation \cite{g-t}. In such
a way, a formal equivalence between descriptions in different
gauges may be established. As we have seen from the examples in
Sect.II and Sect.III the time-dependent gauges in RI theories
may also be connected by means of canonical trransformations
(such a possibility cerrtainly follows from generral theorems
\cite{g-t}). However, such transformations necessary depend on
time (space-time variables). Thus, in this case a formal
possibility to connect different gauges does not mean their
literal physical equivalence. The canonical transformations in
such a case establish only a relation between descriptions of
one and the same system in different RF.

\setcounter{equation}{0}
\section{RI in general and the zero Hamiltonian phenomenon}

Above we have considered several examples of RI systems.
The explicit form of the corresponding GT
depends on the structure of the theory (compare
(\ref{II2}) and (\ref{II3}) ).
At the same time, in all known examples the total Hamiltonian
vanishes on the constraint surface of the theory.
Is it possible to discover some specific structure of
RST in
general and a relation of the latter
with the zero-Hamiltonian phenomenon?
Below we are going to discuss this problem and
present such a relation.

Let us have a theory with finite number of degrees of freedom,
which is described by an action ( $q = q^a, ~ a=1,...,D$
are generalized coordinates  and $t$ is time),
\be
S = \int L(q, \dot{q}, t) dt~.
\ee
Consider a
 transformation in the space of trajectories $q^a(t)$,
\be
q^a(t) \rightarrow q'^a(t) = G^a_t(q)~,                      \label{g1}
\ee
where $G^a_t(q)$ are some  functionals on $q^a(t)$,
depending parametrically on time.
We will call (\ref{g1}) a symmetry transformation (ST) of the theory
if the Lagrangian function $L(q,\dot{q}, t)$ is changed
under such a transformation only by a
total derivative of some function,
\be
L'(q, \dot{q}, t) = L(G_t(q), \dot{G}_t(q), t) =
L(q, \dot{q}, t) + \frac{dF}{dt}~.                           \label{g2}
\ee
One can see that the Lagrangians $L(q,\dot{q},t)$
and $L'(q,\dot{q},t)$ have the same extremals. That can be regarded
as an argument in favor of the proposed definition of the ST.

The ST can be discrete, continuous global
and gauge ones.
Continuous global ST are parametrized by a set of parameters
$\e_{\a}, ~\a = 1,...,r$.
It is convenient to define the point $\e_{\a}=0 $
as the one that corresponds to the identical transformation.
In this case (\ref{g1}) can be presented in the form
\be
q'^a(t) = G^a_t(q|\e), \hskip 5mm   G^a_t(q|0) = q^a(t), \label{g4}
\ee
where the $\e$-dependence is indicated explicitly.
The infinitesimal form of a global continuous ST is:
\be
q'^a(t) =q^a(t) + \d q^a(t),                                  \label{g5}
\hskip 5mm
\d q^a(t) = \r^a_{\a}(t) \e_{\a}, \hskip 5mm
\r^a_{\a}(t) = \left.
\frac{\part G^a_t(q|\e)}{\part \e_{\a}}\right|_{\e=0},
\ee
where  $\r^a_{\a}(t)$ are 
the generators
of the transformations.
Continuous ST are GT (or local ST)
if they are  parametrized by some arbitrary functions on time
(or in the case of field theories by functions of space-time
variables).
They can be presented in the form (\ref{g4}) where,
however, $G^a_t(q|\e)$ may depend not only on $\e$ but on its derivatives
over time.  In this case 
\be
\d q^a(t) = \int R^a_{\a}(t,t') \e_{\a}(t') dt', \hskip 5mm R^a_{\a}(t,t') =
\left.\frac{\d G_t^a(q|\e)}{\d \e_{\a}(t')}\right|_{\e =0}~.
\label{g6}
\ee
As it was demonstrated in \cite{g-t} the generators
$R^a_{\a}(t,t')$ are local in
time (in the case of ordinary bosonic variables) i.e. they have
the following structure
\be
R^a_{\a}(t,t') = \sum^M_{k=0} \r^a_{\a(k)} (t)
\part_t^k \d (t-t')~,                                         \label{g7}
\ee
where $M$ is  finite.
Thus, one can write in this case:
\be
\d q^a(t) = \sum^M_{k=0} \r^a_{\a(k)} (t) \e^{(k)}_{\a}(t),
\hskip 5mm \e^{(k)}_{\a}(t) = \frac{d^k \e_a(t)}{dt^k}~. \label{g8}
\ee

The presence of $r$-parametrical continuous
global ST indicates that there exist $r$ conserved charges.
Indeed, in this case $\d L =  \frac{d }{dt} \d F $,
which is an infinitesimal form of (\ref{g2}).
The variations $\d L$ and $\d F$ can be presented as follows:
\be
\d L = \frac{\d S}{ \d q^a} \d q^a + \frac{d}{dt}
\left( \frac{\part L}{\part \dot{q}} \d q^a \right)
= \left[ \frac{\d S}{\d q^a} \r^a_{\a} +
\frac{d}{dt} \left(\frac{\part L}{\part \dot{q}^a} \r^a_{\a}
\right) \right] \e_{\a},    \hskip 5mm
\d F = f_{\a} \e_{\a},
                               \label{g9}
\ee
where \[
 \frac{\d S}{\d q^a} =
\frac{\part L}{\part q^a} - \frac{d}{dt}
\left( \frac{\part L}{\part \dot{q}^a} \right)~,
\]
so that $ \frac{\d S}{\d q^a} =0$ are the Euler-Lagrange equations
of motion.
Thus, we get
\be
\frac{d Q_{\a}}{dt} = - \r^a_{\a} \frac{\d S}{\d q^a},
\hskip 5mm Q_{\a} = \frac{\part L}{\part \dot{q}^a} \r^a_{\a}
-f_{\a}~,                                               \label{g10}
\ee
and therefore $Q_{\a}$ are the above mentioned conserved charges.
An analogous statement is valid for GT as well.
Moreover, in this case one can make some conclusions
about the structure of the corresponding conserved charges.
Below we are going to formulate and prove some statements, which are
useful for our purposes.

Let an action obey a gauge ST.
In the infinitesimal form that results in the condition
\be
\d L = \frac{\d S}{\d q^a} \d q^a
+ \frac{d}{dt} \left( \frac{\part L}{\part \dot{q}^a}
\d q^a \right) =
\frac{d}{dt} \d F,                                    \label{g11}
\ee
where $\d q^a$ are given by eq.(\ref{g8}) and
$\d F$ is a function.
Similarly to the derivation  (\ref{g9}), (\ref{g10}) that
implies the conservation law
\be
\frac{ d Q}{dt} = - \frac{\d S}{\d q^a} \d q^a,
\hskip 5mm
Q = \left(\frac{\part L}{\part \dot{q}^a} \d q^a - \d F \right)~.
\label{g12}
\ee
The conserved charge $Q$ may be presented in the form:
\be
Q = \sum^{M'}_{k=0} Q_{\a(k)}(t) \e_{\a}^{(k)}(t).   \label{g13}
\ee
Substituting (\ref{g8}) and (\ref{g13}) into
(\ref{g12}) we get
\be
\sum^{M'}_{k=0} \left[ \dot{Q}_{\a(k)} \e_{\a}^{(k)}(t)
+ Q_{\a(k)} \e_{\a}^{(k+1)}(t)\right] =
- \frac{\d S}{\d q^a} \sum^M_{k=0} \r^a_{\a(k)}  \e_{\a}^{(k)}(t)~.
                                                             \label{g14}
\ee
It is clear that $M'= M -1$.
Due to the arbitrariness of  $ \e_{\a}(t)$,
one can consider the derivatives $ \e_{\a}^{(k)}(t)$ as independent
arbitrary functions and compare the terms on the left and right
hand side of (\ref{g14}) with the same  $\e_{\a}^{(k)}(t)$.
Thus one gets:
\begin{eqnarray}
&&Q_{\a(M-1)} = - \frac{\d S}{\d q^a} \r^a_{\a(M)}~,
\;\;\dot{Q}_{\a(M-1)}+  Q_{\a(M-2)} = - \frac{\d S}{\d q^a}
\r^a_{\a(M-1)}~,\hskip 5mm ...~, \nonumber \\
&&\dot{Q}_{\a(k)} +  Q_{\a(k-1)} = - \frac{\d S}{\d q^a}
\r^a_{\a(k)}~ , \hskip 5mm ...~.                         \label{g14r}
\end{eqnarray}
It follows from the system (\ref{g14r}) that
\be
Q_{\a(k)} = \L^a_{\a(k)} \frac{\d S}{\d q^a}~, \hskip 5mm
{\mbox or}\hskip 5mm
Q = \L^a \frac{\d S}{\d q^a},\hskip 5mm  \L^a = \sum^{M-1}_{k=0}
\e_{\a}^{(k)}(t)\L^a_{\a(k)}~,
\label{g15}
\ee
where $ \L^a_{\a(k)}$ contain operators of the
differentiation in time up to the order $(M-k-1)$.
Thus, one may say that\footnote{This statement was, in fact, known to
Noether \cite{Noether}}:

{\em The conserved charge (\ref{g13}) which corresponds to
any GT and its components $Q_{\a(k)}$
vanish on the equations of motion.}

Let a global ST be the reduction of a
GT to constant values of the
parameters $\e_{\a}(t)$.
In this case the generators $\r^a_{\a}(t)$ from
equation (\ref{g5}) are just $\r^a_{\a(0)}(t)$ from equation
(\ref{g8}), and therefore
$
\d q^a(t) = \r^a_{\a(0)}(t) \e_{\a}~.                \label{g17}
$
The corresponding conserved charges $Q_{\a}$ from
(\ref{g10}) coincide with $Q_{\a(0)}$ from (\ref{g13})
and vanish on the equation of
motion according to (\ref{g15}). An inverse statement is also valid, namely:

{\em If some global continuous
ST of an action}~,
$
\d q^a(t) = \r^a(t) \e~, 
$
{\em generates a conserved charge, which vanishes
on the equation of motion, then this action
obeys a gauge symmetry.}

\noindent Let us prove this. 
Similar to eq.(\ref{g9}) one can get:
\be
\frac{\part L}{\part q^a} \r^a +
\frac{\part L}{\part \dot{q}^a} \dot{\r}^a
= \frac{d}{dt} f ~.                                   \label{g20}
\ee
We can use this equation to write the following
relation:
\be
\frac{\part L}{\part q^a} \r^a \e(t) +
\frac{\part L}{\part \dot{q}^a} \frac{d}{dt}
\left[ \r^a \e(t) \right] =
\frac{d}{dt} \left[f \e(t)\right] + \dot{\e}(t) Q_{(0)}~, \label{g21}
\ee
where $ Q_{(0)} = \frac{\part L}{\part \dot{q}^a}\r^a -f$
is the conserved charge related to the global continuous ST
(see (\ref{g10}), and $\e(t)$ an arbitrary function
of $t$.
Let this charge vanish on the equations of motion, that is
\be
Q_{(0)} = \L_{(0)}^a \frac{\d S}{\d q^a}~,             \label{g22}
\ee
where $\L_{(0)}^a$ may contain operators of differentiation
with respect to time up to a finite order.
Thus, the last term in the right hand side of
(\ref{g21}) has the form
$
\dot{\e}(t) \L_{(0)}^a \frac{\d S}{\d q^a}$.
One can always write this term in the different form:
\be
\dot{\e}(t) \L_{(0)}^a \frac{\d S}{\d q^a}  =
- \frac{\d S}{\d q^a} \L^a \dot{\e}(t) + \frac{d \v}{dt}~,\label{g24}
\ee
where $\L^a$ is an operator  symmetric to  $\L^a_{(0)}$,
and $\v$ some function.
On the other hand
\be
\frac{\d S}{\d q^a} \L^a \dot{\e}(t) =
\frac{\part  L}{\part q^a} \L^a \dot{\e}(t) +
\frac{\part  L}{\part \dot{q}^a} \frac{d}{dt} \left[
\L^a \dot{\e}(t) \right] -
 \frac{d}{dt} \left[\frac{\part  L}{\part \dot{q}^a} \L^a
 \dot{\e}(t) \right]~.                                      \label{g25}
\ee
Gathering (\ref{g21}), (\ref{g24}) and (\ref{g25}) we get
\[
\d L = \frac{\part  L}{\part q^a} \d q^a(t) +
\frac{\part  L}{\part \dot{q}^a} \d \dot{q}^a(t)
= \frac{d}{dt} \left[ f \e(t) + \v +
\frac{\part  L}{\part \dot{q}^a} \L^a \dot{\e}(t) \right]~,
\]
where $\d q^a(t)$ is a GT,\be
\d q^a(t) = \r^a \e(t) + \L^a \dot{\e}(t)~.  \label{g26}
\ee
Based on the two statements proved above we may
 define what can be called
reparametri\-zation ST
in general.
To this end let us first discover what is a global
representative of such a symmetry.
One can remember that in all known examples the existence
of the reparametrization invariance leads to the zero-Hamiltonian
phenomenon.
More exactly, the total Hamiltonian \cite{ab1,g-t} appears to
be proportional to  constraints of the theory, or it  vanishes
on the equations of motion.
Such a Hamiltonian can be derived from the expression
for the Lagrangian energy,
if one replaces there all the primary-expressible velocities
as functions on phase space variables and denotes the primary
unexpressible velocities by $\l$, which play then the role
of Lagrange multipliers.
Thus, in this case
one can write
\be
{\cal E} =\frac{\part L}{\part \dot{q}^a}\dot{q}^a - L=
 \L_{(0)}^a \frac{\d S}{\d q^a}~.                     \label{g28}
\ee
Another observation is that in all known 
examples, where RI takes place, the corresponding Lagrangians do not
depend explicitly on time.
Thus, we have the conservation law:
\be
\frac{ d{\cal E}}{dt} 
= - \dot{q}^a \frac{\d S}{\d q^a}~.     \label{g29}
\ee
On the other hand, one can interpret energy ${\cal E}$ as a conserved
charge related to the global ST,
which are translations in time, $q^a(t) \rightarrow
q^a(t + \e)$ or in the infinitesimal form:
\be
\d q^a(t) = \dot{q}^a(t) \e~.                                \label{g30}
\ee
Indeed, in this case
\be
\d L =  \frac{\part L}{\part q^a} \dot{q}^a \e +
 \frac{\part L}{\part \dot{q}^a} \ddot{q}^a \e
= \e \frac{dL}{dt}~,                                          \label{g31}
\ee
so that (\ref{g30}) is a symmetry and, at the same time,
(\ref{g29}) follows also from (\ref{g31}).
Taking all said into account it is natural to regard
 translations in time as global representatives of the
reparametrization GT.
Then, one can define the latter GT
as a possible extension of the translations in time to  GT,
in the manner which was used in the proof
of the inverse statement.
Thus,  such GT
have the form (\ref{g26}) with $\r^a = \dot{q}^a$,
\be
\d q^a(t) = \dot{q}^a(t) \e(t) + \L^a \dot{\e}(t)~,     \label{g32}
\ee
where the operators $\L^a$ are defined by the explicit form
of the Lagrangian of the theory (see for example the transformations
(\ref{II2}) and (\ref{II4})).

Considering the above finite-dimensional case, we have seen
that the conserved charge $Q$ (\ref{g13}) related to any GT
and all its components $Q_{\a(\k)}$ vanish on the equations of motion.
In particular, the components $Q_{\a(0)}$, which are
the conserved charges related to the corresponding global ST
(global representatives of the GT), with
$\e_{\a}(t) = \e_{\a} = const$, also vanish on the equations of
motion.
However, such a conclusion may be wrong in the case of field theory.
As an example, let us take electrodynamics coupled to a scalar field
$\v(x)$,
\be
S = \int {\cal L} d^{D+1} x, \hskip 5mm
{\cal L} = - \frac{1}{4} F_{\m\n}F^{\m\n} +
(\part_{\m} + i e A_{\m}) \v^{\dagger}
(\part^{\m} - i e A^{\m}) \v
 - V(\v^{\dagger} \v)~.
\ee
The conserved charge, (an analog of (\ref{g10})), related to GT
$ \d A_{\m}(x) = \part_{\m} \e(x),~ \d\v(x) = i e \v(x) \e(x),~
\d \v^{\dagger}(x) = -i e \v^{\dagger}(x) \e(x),$
where $\e(x)$ are parameters of the GT, is
\be
Q = \int \left( \frac{\part L}{\part \dot{A}_{\m}} \d A_{\m}
+ \frac{\part L}{\part \dot{\v}} \d \v+
\frac{\part L}{\part \dot{\v}^{\dagger}} \d \v^{\dagger} \right)
d^D x =
\int \left[ F_{0k} \part_k \e(x) - j_0 \e(x) \right] d^D x~,
\ee
\be
j_0 = \v^{\dagger} (\partial_0 - i e A_0) \v
- \v ( \partial_0 + i e A_0) \v^{\dagger}~.
\ee
This expression can be transformed on the equation of motion
$\part_k F_{0k} + i e j_0 =0$,
to the following form
\be
Q = \int  \part_k \left[ F_{0k} \e(x) \right] d^D x~. \label{haromcs}
\ee
In the case of GT with $\e(x)$ decreasing rapidly enough
in the limit $\mid {\bf x} \mid \rightarrow  \infty$,
 the charge (\ref{haromcs}) is zero.
In the case of global ST with $\e(t) = \e = const$, we have
\be
Q = \e \int \part_k F_{0k} d^D x = - i e \e \int j_0 d^D x~.
\label{negycs}
\ee
This expression may differ from zero.
In the Coulomb phase $F_{0k}$ behaves at large $r$ as $r^{-(D-1)}$,
so that the integral in (\ref{negycs}) is proportional to the
total electrical charge of the system, which is in general not zero.
However, if a spontaneous symmetry braking takes place (Higgs phase)
the vector field becomes massive and $F_{0k}$ decreases
exponentially,
resulting in $Q=0$.
(The total charge of any state is zero.)

One meets a similar situation in the theory of gravity.
Let us select the action of the gravitational field of the
form, which was first proposed by Dirac \cite{dirac2}
(for a detailed treatment see \cite{LL,g-t}),
\be
S = \int L d^4 x,
\hskip 5mm L = A + \part_i q^i~,
\ee
where
\[
A = \sqrt{- g^{00} g_{(3)}} \left[ \frac{z_{ik}}{4}
\left( e^{il} e^{km} - e^{ik} e^{lm} \right) z_{lm} -
- \frac{R_{(3)}}{g^{00}}\right]\;,
\hskip 3mm
g_{(3)} =  \mid g_{ik}  \mid\;, \hskip 3mm
e^{ik} g_{kl} = \d^i_l\;,
\]
\[
z_{ik} = \dot{g}_{ik} - g_{0i,k} - g_{0i,k} + 2 \g^l_{ik} g_{0l},
\hskip 5mm
q^i = \sqrt{-g_{(3)}} g_{lm,k} \left( e^{il} e^{km} - e^{ik} e^{lm}
\right),
\]
and $\g^l_{ik}$, $R_{(3)}$ are the Christoffel symbols and the
scalar curvature constructed for the three-dimensional metric
$g_{ik}$.
This action
 is equivalent to the
Einstein-Hilbert one  under certain assumptions about the
global structure of the theory.
The Lagrangian $L$ contains neither higher (second) order
derivatives of the metric, nor velocities $\dot{g}_{0\m}$.
The variation of $L$ under the GT (\ref{deltag}) has the form
$
\d L = \part_{\m} [ L \e^{\m} (x) ].
$
The corresponding conserved charge is
\be
Q = \int \left( \frac{\part L}{\part \dot{g}_{ik}} \d g_{ik}
- L \e^0 \right) d^3 x.                \label{+}
\ee
If $\e^{\m} (x) \rightarrow 0$ when
$\mid {\bf x} \mid \rightarrow \infty$, then one can see that it vanishes on
the equations of motion. For example, if $\e^i(x)\equiv 0$
\be
Q = \int \e^0\left[g_{\mu\nu}\frac{\d S}{\d g_{\mu\nu}} +
g_{00}\frac{\d S}{\d g_{00}}\right] d^3 x~.
\ee
In the case of $\e_0(x) = \e_0 = const, ~ \e_i(x) \equiv 0,$ the charge
(\ref{+}) is proportional to the total energy and has the form:  \be Q = -
\e_0 \int \part_i q^i d^3 x~.  \label{++} \ee The integral on the right hand
side of (\ref{++}) is generally non-zero.  In particular, in an
asymptotically flat space \cite{fadeev} for the system with the total mass
$M$ \be g_{ik} = - \d^i_k \left( 1+ \frac{M}{8 \p r} \right) + O\left(
\frac{1}{r^2} \right).  \ee Then $Q = \e_0 M$ is not zero.  One can remark,
considering for example the theory of gravity, that in spite of the fact that
four-dimensional divergency terms in the Lagrangian do not affect the form of
the equations of motion, they can affect the form of the corresponding
conserved charges.  That may serve as an additional argument in favor of a
certain form of the selected Lagrangian.

{\bf Acknowledgement}
The authors are thankful  to foundations
FAPESP (F\"ul\"op), CNPq (Gitman) and to INTAS 96-0308, FRBR 96-02-17314
(Tyutin).

 \end{document}